\documentclass{aastex631}

\pdfoutput=1

\usepackage{amsmath}

\newcommand{\mcgillphysics}{Department of Physics, McGill University, 3600 rue University, Montr\'eal, QC H3A 2T8, Canada}
\newcommand{\msi}{McGill Space Institute, McGill University, 3550 rue University, Montr\'eal, QC H3A 2A7, Canada}

\newcommand{\wvugws}{Center for Gravitational Waves and Cosmology, West Virginia University, Chestnut Ridge Research Building, Morgantown, WV 26505, USA}
\newcommand{\wvucs}{Lane Department of Computer Science and Electrical Engineering, 1220 Evansdale Drive, PO Box 6109  Morgantown, WV 26506, USA}

\newcommand{\cita}{Canadian Institute for Theoretical Astrophysics, 60 St. George Street, Toronto, ON M5S 3H8, Canada}
\newcommand{\dunlapinstitute}{Dunlap Institute for Astronomy \& Astrophysics, University of Toronto, 50 St. George Street, Toronto, ON M5S 3H4, Canada}
\newcommand{\dunlapdep}{David A. Dunlap Department of Astronomy \& Astrophysics, University of Toronto, 50 St. George Street, Toronto, ON M5S 3H4, Canada}
\newcommand{\nrao}{National Radio Astronomy Observatory, 520 Edgemont Rd, Charlottesville, VA 22903, USA}
\newcommand{\haa}{National Research Council Canada, Herzberg Astronomy and Astrophysics Research Centre, Dominion Radio Astrophysical Observatory, PO Box 248, Penticton, British Columbia, V2A 6J9 Canada}
\newcommand{\mitkavli}{MIT Kavli Institute for Astrophysics and Space Research, Massachusetts Institute of Technology, 77 Massachusetts Ave, Cambridge, MA 02139, USA}
\newcommand{\mitphysics}{Department of Physics, Massachusetts Institute of Technology, 77 Massachusetts Ave, Cambridge, MA 02139, USA}
\newcommand{\ubc}{Dept. of Physics and Astronomy, 6224 Agricultural Road, Vancouver, BC V6T 1Z1 Canada}
\newcommand{\sidrat}{Sidrat Research, PO Box 73527 RPO Wychwood, Toronto, Ontario, M6C 4A7, Canada}
\newcommand{\perimeter}{Perimeter Institute for Theoretical Physics, 31 Caroline Street N, Waterloo ON N2L 2Y5 Canada}
\newcommand{\waterloo}{Department of Physics and Astronomy, University of Waterloo, Waterloo, ON N2L 3G1, Canada}
\newcommand{\tata}{Department of Astronomy and Astrophysics, Tata Institute of Fundamental Research, Mumbai, 400005, India}
\newcommand{\ncra}{National Centre for Radio Astrophysics, Post Bag 3, Ganeshkhind, Pune, 411007, India}

\begin{document}
\title{Fast Radio Burst Morphology in the First CHIME/FRB Catalog}
\shorttitle{FRB Morphology in the First CHIME/FRB Catalog}

\correspondingauthor{Ziggy Pleunis}
\email{ziggy.pleunis@physics.mcgill.ca}

\author[0000-0002-4795-697X]{Ziggy~Pleunis}
\affiliation{\mcgillphysics}
\affiliation{\msi}

\author[0000-0003-1884-348X]{Deborah~C.~Good}
\affiliation{\ubc}

\author[0000-0001-9345-0307]{Victoria~M.~Kaspi}
\affiliation{\mcgillphysics}
\affiliation{\msi}

\author[0000-0001-7348-6900]{Ryan~Mckinven}
\affiliation{\dunlapinstitute}
\affiliation{\dunlapdep}

\author[0000-0001-5799-9714]{Scott~M.~Ransom}
\affiliation{\nrao}

\author[0000-0002-7374-7119]{Paul~Scholz}
\affiliation{\dunlapinstitute}

\author[0000-0003-3772-2798]{Kevin~Bandura}
\affiliation{\wvucs}
\affiliation{\wvugws}

\author[0000-0002-3615-3514]{Mohit~Bhardwaj}
\affiliation{\mcgillphysics}
\affiliation{\msi}

\author[0000-0001-8537-9299]{P.~J.~Boyle}
\affiliation{\mcgillphysics}
\affiliation{\msi}

\author[0000-0002-1800-8233]{Charanjot~Brar}
\affiliation{\mcgillphysics}
\affiliation{\msi}

\author[0000-0003-2047-5276]{Tomas~Cassanelli}
\affiliation{\dunlapinstitute}
\affiliation{\dunlapdep}

\author[0000-0002-3426-7606]{Pragya~Chawla}
\affiliation{\mcgillphysics}
\affiliation{\msi}

\author[0000-0003-4098-5222]{Fengqiu~(Adam)~Dong}
\affiliation{\ubc}

\author[0000-0001-8384-5049]{Emmanuel~Fonseca}
\affiliation{\mcgillphysics}
\affiliation{\msi}

\author[0000-0002-3382-9558]{B.~M.~Gaensler}
\affiliation{\dunlapinstitute}
\affiliation{\dunlapdep}

\author[0000-0003-3059-6223]{Alexander~Josephy}
\affiliation{\mcgillphysics}
\affiliation{\msi}

\author[0000-0003-4810-7803]{Jane~F.~Kaczmarek}
\affiliation{\haa}

\author[0000-0002-4209-7408]{Calvin~Leung}
\affiliation{\mitkavli}
\affiliation{\mitphysics}

\author[0000-0001-7453-4273]{Hsiu-Hsien~Lin}
\affiliation{\cita}

\author[0000-0002-4279-6946]{Kiyoshi~W.~Masui}
\affiliation{\mitkavli}
\affiliation{\mitphysics}

\author[0000-0002-0772-9326]{Juan~Mena-Parra}
\affiliation{\mitkavli}

\author[0000-0002-2551-7554]{Daniele~Michilli}
\affiliation{\mcgillphysics}
\affiliation{\msi}

\author[0000-0002-3616-5160]{Cherry~Ng}
\affiliation{\dunlapinstitute}

\author[0000-0003-3367-1073]{Chitrang~Patel}
\affiliation{\dunlapinstitute}
\affiliation{\mcgillphysics}

\author[0000-0001-7694-6650]{Masoud~Rafiei-Ravandi}
\affiliation{\perimeter}
\affiliation{\waterloo}

\author[0000-0003-1842-6096]{Mubdi~Rahman}
\affiliation{\dunlapinstitute}
\affiliation{\sidrat}

\author[0000-0001-5504-229X]{Pranav~Sanghavi}
\affiliation{\wvucs}
\affiliation{\wvugws}

\author[0000-0002-6823-2073]{Kaitlyn~Shin}
\affiliation{\mitkavli}
\affiliation{\mitphysics}

\author[0000-0002-2088-3125]{Kendrick~M.~Smith}
\affiliation{\perimeter}

\author[0000-0001-9784-8670]{Ingrid~H.~Stairs}
\affiliation{\ubc}

\author[0000-0003-2548-2926]{Shriharsh~P.~Tendulkar}
\affiliation{\tata}
\affiliation{\ncra}

\begin{abstract}
We present a synthesis of fast radio burst (FRB) morphology (the change in  flux as a function of time and frequency) as detected in the 400--800\,MHz octave by the FRB project on the Canadian Hydrogen Intensity Mapping Experiment (CHIME/FRB), using events from the first CHIME/FRB catalog. The catalog consists of 61 bursts from 18 repeating sources, plus 474 one-off FRBs, detected between 2018 July 25 and 2019 July 2. We identify four observed archetypes of burst morphology (``simple broadband,'' ``simple narrowband,'' ``temporally complex'' and ``downward drifting'') and describe relevant instrumental biases that are essential for interpreting the observed morphologies. Using the catalog properties of the FRBs, we confirm that bursts from repeating sources, on average, have larger widths and we show, for the first time, that bursts from repeating sources, on average, are narrower in bandwidth. This difference could be due to a beaming or propagation effects, or it could be intrinsic to the populations. We discuss potential implications of these morphological differences for using FRBs as astrophysical tools.
\end{abstract}

\keywords{Radio transient sources (2008); High energy astrophysics (739)}

\section{Introduction}

Fast radio bursts \citep[FRBs;][]{lbm+07} are $\mu$s--ms dispersed impulses of radio waves that are observed out to cosmological distances \citep[see][for reviews of the phenomenon]{cc19, phl19}. Despite a growing number of detected bursts, host galaxy associations, and theoretical efforts, the origins of FRBs remain elusive. Progenitor models involving cataclysmic events were disfavored by the initial discovery of repeat bursts from an FRB \citep{ssh+16a} and the establishment of repeating sources of FRBs as a class \citep{abb+19b,abb+19c}. Many progenitor models include compact objects, with the FRB emission originating inside or outside of the magnetosphere, or in shocks in the circumstellar material\footnote{\url{https://frbtheorycat.org/} provides a living review of FRB models.} \citep{pww+19}. The recent discovery of a $\sim$MJy radio burst from Galactic magnetar SGR 1935+2154 \citep{abb+20,brb+20} has shown that magnetars are viable candidates as the sources of at least some fraction of FRBs.

It remains an open question whether all FRBs repeat. So far, it is only possible to decide if any one FRB source is repeating by spending considerable telescope time on follow-up observations \citep{cp18}. Even after many hundreds of hours of nondetections, only upper limits on repeat rates can be placed and the matter is further complicated by observed orders of magnitude differences in fluence between the initial discovery and repeat bursts of some sources \citep{kso+19}. Definite proof of the nonrepeating nature of an FRB will likely only be provided when an FRB is observed to be associated with a cataclysmic counterpart, such as a supernova or gamma-ray burst. By means of repetition statistics one can attempt to answer whether all sources repeat \citep{csr+19,jof+20a,jof+20b}. An alternative means of classifying an FRB as a one-off event or a repeater burst, even if only probabilisitically, would therefore be useful. This will aid the selection of the most promising targets for follow-up observations and could help uncover the variety of FRBs. The classification of an FRB could, for example, be done by way of its spectrotemporal and polarimetric properties or its (position in a) host galaxy.

Burst morphology (the change in flux as a function of time and frequency) is a powerful proxy for understanding burst emission and propagation. FRBs were initially detected in $\sim$ms time resolution, total intensity data, and morphology was often discussed in terms of burst structure in band-averaged time series instead of in dynamic spectra of FRBs.\footnote{For example, no dynamic spectrum was presented with the first evidence of a multicomponent burst \citep{cpk+16}; but see \citet{rav19b}.} Population studies of FRB morphology have become feasible with the advent of large (uniform) samples of FRBs. Moreover, with many instruments now capable of saving complex voltage data following the real time detection of an FRB, burst morphology can be studied in unprecedented detail for smaller samples of individual bursts as well: at $\mu$s time resolution and with full polarization information.

Burst structures on timescales of $\sim30\,\mu$s have been observed for both one-off events and repeater bursts \citep{ffb18,msh+18}. \citet{nhk+21} detect burst timescales spanning three orders of magnitude for FRB 20180916B, with the briefest bursts lasting only 3--4\,$\mu$s. This constrains the size of the emission region to be less than about 1\,km. Comparing Arecibo detections of bursts from FRB 20121102A with the Parkes sample of one-off events revealed an apparent difference in bursts widths \citep{sbh+17} that was later confirmed by comparing repeater bursts and one-off events discovered by the CHIME/FRB project \citep{abb+19c,fab+20}. Furthermore, for FRBs 20121102A and 20180916B, burst widths tend to be narrower at higher frequencies \citep{gsp+18,nhk+21}. At $\mu$s time resolution, a polychotomy of classes seems to appear, between broad singlecomponent one-off events, narrow and multicomponent one-off events, and repeater bursts that exhibit downward-drifting chromatic subbursts \citep{dds+20}.

FRBs have now been detected from 110\,MHz to 8\,GHz \citep{gsp+18,pcl+21,pmb+21}. Bright apparent nonrepeaters detected by the Australian Square Kilometre Array Pathfinder (ASKAP) show knotty, strongly modulated spectra with power concentrated in patches only a few MHz wide \citep{smb+18}. As an ensemble, 23 ASKAP FRBs have an average spectrum that is fitted well by a power-law model with spectral index $-1.4$ or $-1.6$ reminiscent of average radio pulsar spectra \citep{msb+19}. Many bursts from repeaters, on the other hand, show Gaussian-like emission bandwidths of 100--200\,MHz in the CHIME band \citep[400--800\,MHz; e.g.,][]{abb+19c}, and up to a GHz at 4--8\,GHz \citep{gsp+18}, at varying central frequencies.

Subbursts that drift downward in frequency\footnote{Colloquially known as the ``sad trombone'' effect.} were first observed in FRB~20121102A \citep{gsp+18} and studied in great detail by \citet{hss+19}. The phenomenon was also seen in bursts from the second detected repeating source of FRBs \citep{abb+19b} and now seems ubiquitous among repeaters \citep{abb+19c,fab+20,dds+20}. It has not been observed in all bursts or for all repeating sources, but this might be due to the structures being unresolved with limited time resolution or sensitivity \citep{gms+19}. Complex structure has also been observed in FRBs that have not yet been observed to repeat \citep[see, e.g.,][]{ffb18,cms+20}. Wide burst envelopes of up to tens of ms are often comprised of multiple spectrotemporal components that can be relatively wide in time in the ``wings'' of the burst envelope and sharp and narrow at the center of emission. The exact appearance of the structures depends on the choice of metric for dispersion optimization, but aligning subbursts assuming all frequencies in one component were emitted at the source at the same time leads to the least scatter in dispersion measure \citep[DM;][]{hss+19}.

Drift rates, d$\nu$/d$t$ (assumed to be linear over the band in which they occur), are measured to be a few to tens of MHz ms$^{-1}$ in the CHIME band for all sources where the drift is apparent. FRBs 20121102A and 20180916B so far are the only source for which drift has been measured at higher frequencies, and the drift evolves to about GHz ms$^{-1}$ at 6.5\,GHz \citep{hss+19} with an apparent linear trend amidst large scatter \citep{jcf+19,csa+20,pcl+21}. Recent wideband and sensitive observations have shown that subsequent subbursts spaced $\lesssim 50$ ms can exhibit order-of-magnitude differences in brightness \citep{cab+20,csa+20}.

In summary, FRBs show a variety of morphologies and there are apparent differences between the appearance of one-off events and repeater bursts. Any comparisons done so far, however, are between detections with different telescopes, with different selection functions, or with a limited sample size. A more systematic study on a large sample of FRBs discovered with the same selection function is now possible using the first CHIME/FRB catalog, and is presented here. In \S\ref{sec:obs} we summarize the creation of the first CHIME/FRB catalog, and we discuss relevant biases that affect observed burst morphologies. In \S\ref{sec:archetypes} we categorize the catalog into four observed morphological archetypes of FRBs. We compare one-off events and repeater bursts statistically in \S\ref{sec:versus}. We present subburst separations in \S\ref{sec:subbursts}. We discuss our findings in \S\ref{sec:discussion} and conclude in \S\ref{sec:conclusion}.

\section{Observations and Analysis}
\label{sec:obs}

The first CHIME/FRB catalog \citep[][hereafter ``Catalog 1'']{chimefrbcatalog1} presents all FRBs seen by the experiment between 2018 July 25 and 2019 July 2: 61 bursts from 18 repeating sources, plus 474 one-off events.\footnote{The repeater bursts in Catalog 1 have all been published before, in \citet{abb+19c,jcf+19,fab+20}.} Catalog 1 contains exposure and sensitivity metrics for all FRB sources during the time frame of the catalog. Events are localized based on beam detection S/N ratios, which typically gives a $\sim$15$^\prime$ precision (for 68\% of Catalog 1), but for the highest S/N events the localization can be as good as a few arcminutes. In Catalog 1, we analyze the intensity data of the beam in which each FRB was detected with the highest S/N. These data have Stokes $I$ total intensity at a 0.98304 ms time resolution and with 16,384 frequency channels over 400\,MHz. Here, we are not correcting the observed distributions of FRBs for the selection function as in Catalog 1. We do exclude three events that were very likely detected in far sidelobes from our analysis (see Catalog 1).

We model all bursts with the least-squares fitting described in Catalog 1, allowing for multiple components. In summary, a burst model is described by one DM and one scattering timescale $\tau_\mathrm{scat}$ with a $\nu^{-4}$ frequency dependence. A burst can consist of multiple components, each with its own intrinsic width and spectrum. The intrinsic widths are defined as the standard deviation $\sigma$ of the Gaussian burst that are subsequently broadened by dispersion smearing and convolution with an exponential scattering tail \citep{mck14}. All burst components are modeled at a higher resolution than the data ($8\times$ in time, $4\times$ in frequency) to model dispersion smearing, before being averaged to the instrumental resolution. Based on simulations, we know that we can conservatively measure intrinsic widths and scattering timescales at 600\,MHz accurately down to 100\,$\mu$s, under the assumption that bursts are comprised of a single component that spans the full receiver bandwidth. For high S/N events we can likely get more precise results, as long as the models used are an accurate description of the data. Beyond S/N, the precision to which burst parameters can be determined depends on the center frequency and bandwidth of the bursts, and the exact dependence will be studied in the future.

CHIME FRBs that are not detected in far sidelobes\footnote{Detections in far sidelobes show a number of narrow knots of emission ($\sim$10\,MHz wide each, separated by $\sim$10 to $\sim$250\,MHz) due to the chromatic beam response in the far sidelobes, see the dynamic spectra of FRBs~20190125B, 20190202B and 20190210D in Catalog 1 for examples. In future work we will discuss how the spacing of these narrow features can be used to localize the bursts.} have observed spectra that either, i.) extend over the full bandwidth or peak at the edge of the band and seem to be best described by a power law, as is common for radio pulsars, or ii.) are band-limited and seem to be best described by a top-hat or Gaussian function. The two extremes can be relatively well described by one continuous power-law function if we take the common radio spectral index $\gamma$ and add an extra ``running'' term $r$ in the exponent\footnote{This is analogous to the parametrization of scalar fluctuations in the primordial power spectrum used in cosmology \citep[see, e.g.,][]{planck2018vi}. There, the power spectrum $\mathcal{P}_\mathcal{R}(k) = A_s (k / k_0)^{n(k)}$ with spectral index $n_s$ and its first derivative with respect to $\ln k$ (the ``running'' of the spectral index) is $n(k) = n_s - 1 + (1/2)(dn_s/d\ln k) \ln(k/k_0)$. Whereas in the cosmological context the spectral index and running of the spectral index relate to the physics of the early Universe, the parametrization of FRB spectra here is as-of-yet purely empirical, and has no physical underpinning.}
\begin{align}
I(\nu) = A (\nu / \nu_0)^{\gamma + r \ln(\nu/\nu_0)},
\label{eq:spectra}
\end{align}
where $I(\nu)$ is the intensity at spectral frequency $\nu$, $A$ the amplitude and $\nu_0$ the pivotal frequency, taken to be 400.1953125\,MHz, the bottom of the CHIME band.

\begin{figure}
    \centering
    \includegraphics{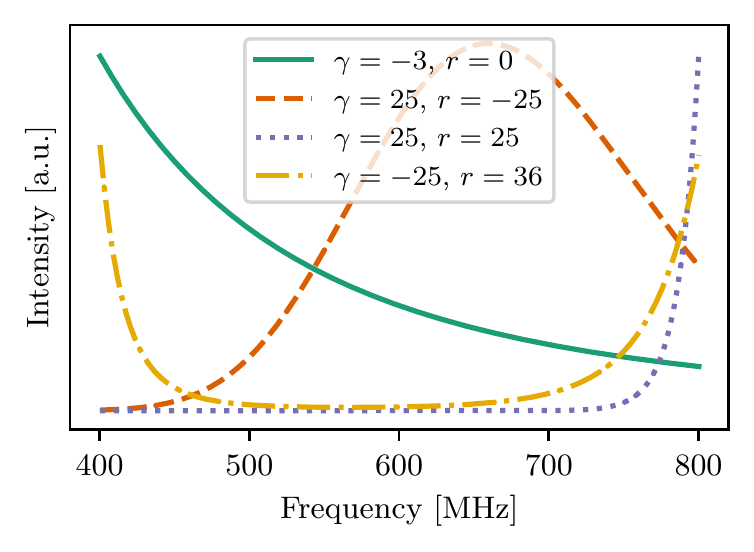}
    \caption{Examples of spectral shapes as parametrized by Equation~\eqref{eq:spectra} for different combinations of spectral index $\gamma$ and running of the spectral index $r$. Where $\gamma$ is negative and $r = 0$ the function is equivalent to a power law peaking at lower frequencies (green solid curve). Similarly, positive $\gamma$ and $r = 0$ lead to a power law peaking at higher frequencies. Positive $\gamma$ with negative $r$ approximates a Gaussian function (orange dashed curve). With positive $\gamma$ and positive $r$ the result is a steeply rising spectrum at the top part of the band (purple dashed curve) and with negative $\gamma$ and positive $r$ the spectrum has asymmetric peaks on either end of the band (yellow dash-dotted curve). The vertical axis is scaled linearly.}
    \label{fig:fitburst_spectra}
\end{figure}

In the limit where $r \rightarrow 0$ this approaches a power-law function. For positive $\gamma$ and negative $r$ this approximates a Gaussian function reasonably well. In the other extremes it gives more unrealistic spectra, with a very steep rise on either end of the band or with a lack of emission in the middle of the band. Example spectra for possible combinations of $\gamma$ and $r$ are depicted in Figure~\ref{fig:fitburst_spectra}. Note that the fitted spectral indices from using equation~\eqref{eq:spectra} cannot be directly compared to the spectral indices from fits using a power-law function without a running term (where $r = 0$). The bandwidth and peak frequency can be found directly from evaluating the functional form. Alternatively, the peak frequency can be found from completing the square of the exponent in the Gaussian form of Equation~\eqref{eq:spectra}, which results in $\nu_\mathrm{vertex}=\nu_0\exp\left[-\gamma/\left(2r\right)\right]$.

On top of the burst modeling, we measure burst durations by boxcar convolution with the dedispersed and frequency-averaged time series. The measurement of burst duration makes no assumptions about scattering and encapsulates multiple subbursts in the case of multicomponent bursts. The width of the boxcar that leads to highest S/N is the duration of the burst. The burst bandwidths are derived from the time-averaged burst models, and they are defined at the full width at tenth maximum (FWTM) of the peak and capped at the top and bottom of the CHIME band.

\subsection{Biases}

Instrumental and analysis biases, such as the beam response for off-axis detections and limited time resolution which leads to unresolved bursts, lead to known biases in the interpretation of FRB morphology. We list various examples of bias next.

\subsubsection{Time resolution}

FRBs are $\mu$s--ms transients with many of them comprised of multiple subbursts with sub-ms widths \citep{nhk+21}. At the 0.98304 ms time resolution of the CHIME/FRB intensity data, the sub-ms structure of bursts will be unresolved and a multicomponent FRB might be interpreted as having just one component. Scatter-broadening can also blend multiple intrinsic burst components \citep[e.g., \S4.3 in][]{dds+20}. Moreover, any transient search conducted at $\sim$1 ms time resolution will will have a lower sensitivity for sub-ms FRBs \citep{connor19}.

\subsubsection{Dispersion measure}

DMs measured from data at lower time resolution using incoherent dedispersion (i.e., the data analyzed here) are less precise than DMs measured from high-time resolution complex voltage data, where coherent dedispersion is possible. The DMs measured at lower time resolution are, however, unbiased (i.e., they are not systematically lower or higher) with respect to the DMs measured at higher time resolution; as long as center-of-channel frequencies are used for the dispersion correction and frequency channels are sufficiently narrow such that there is very little DM ``curvature'' (i.e., from $\nu^{-2}$) across the channel. In the case of narrow frequency channels, the best S/N will occur when all of the smeared channels are symmetrically lined up, and the centers of the smeared profiles are defined by the center-of-channel frequencies. CHIME/FRB has 16,384 frequency channels over 400\,MHz and is in the narrow-channel regime. Dedispersion with respect to the bottom or top frequencies of channels will result in a DM bias (of order $10^{-4}$ pc~cm$^{-3}$ for top-of-the-channel dedispersion of CHIME/FRB intensity data, as determined from simulations).

What we measure as the DM of FRBs comprised of downward-drifting subbursts depends on the optimization metric used. Dedispersed to higher DMs, subbursts can superimpose and lead to higher frequency-averaged S/Ns \citep{gsp+18}. When optimized for S/N, DMs from repeaters show greater burst-to-burst variation than when they are optimized for structure \citep[i.e., when subbursts are aligned;][]{hss+19,abb+19c}. When FRBs are modeled with multiple Gaussian components, the DMs tend toward values that maximize structure (as long as the Gaussians are reasonable approximations of the subbursts).

When an FRB is comprised of downward-drifting subbursts that remain unresolved, DMs are biased high \citep{aab+20}. The exact bias depends on the drift rate, subburst separation and instrumental resolution \citep{jcf+19}, but we will not attempt to quantify it here. As an example for CHIME/FRB, we take two bursts from FRB 20180916B \citep{aab+20} that show clear downward-drifting in high-time resolution CHIME/FRB baseband data and downsample those data to the 0.98304 ms resolution of CHIME/FRB intensity data, before optimizing DMs. At high-time resolution these bursts have structure-optimizing DMs of 348.78(2) and 348.82(5) pc~cm$^{-3}$, respectively, but at downsampled time resolution the structure-optiminizing DMs are 349.5(3) and 349.3(1) pc~cm$^{-3}$, where the uncertainties are the standard deviation among trials with different downsampling offsets, showing that DMs can be overestimated by 0.5--1 pc~cm$^{-3}$ in this small sample. We will investigate this bias further in future publications of CHIME/FRB baseband data \citep[see][for an overview of the CHIME/FRB baseband analysis pipeline]{mmm+20}.

\subsubsection{Observed spectra}
\label{sec:beameffects}

\begin{figure}
    \centering
    \includegraphics{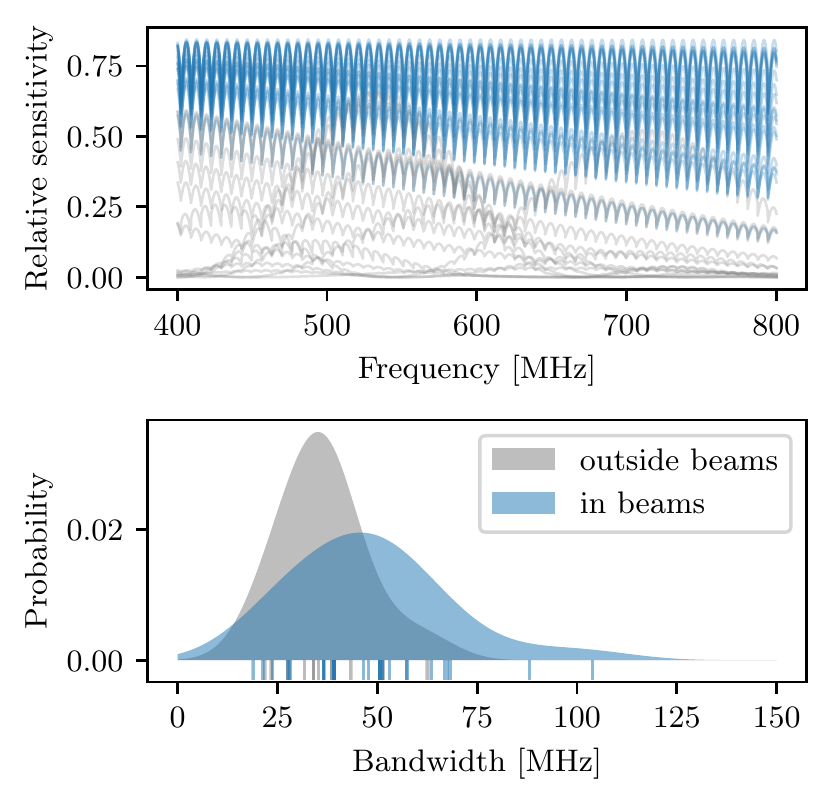}
    \caption{Top: Relative beam sensitivity across the band to bursts from the localized FRB~20180916B at the time of their detection, inside (blue) and outside (gray) the FWHM of the formed beams at 600\,MHz based on the best available beam model, which includes both the primary beam response and the synthesized beams. The rapid variations are due to a ``clamping'' step in the beam forming \citep{nvp+17}. Sensitivity variations due to the polyphase filterbank in the spectrometer are not included here and would appear on a much smaller scale. Bottom: Bandwidth FWTM of CHIME/FRB detections of FRB~20180916B, separated into bursts that were detected inside (30; blue) and outside (16; gray) the formed beams show that burst emission is supressed outside the formed beams. Bandwidths are taken from \citet{abb+19c} and \citet{aab+20}. Both the values (vertical lines) and kernel density estimations are shown (see \S\ref{sec:beameffects}).}
    \label{fig:r3bws}
\end{figure}

There is a bias due to the chromatic reduction in sensitivity if a burst is detected away from the center of the formed beams. Even though the response of all 1,024 beams in CHIME/FRB as a function of frequency and sky position is encapsulated in a beam model, the uncertainty on the sky position of each FRB in Catalog 1 is not precise enough to meaningfully correct for the beam response. When a source of repeating FRBs is sky-localized to high enough precision, as is the case for FRB~20180916B \citep{mnh+20}, it is known which bursts were detected in and out of the beams. Here, as elsewhere, we define the beam edge as the FWHM at 600\,MHz of the beams. Comparing the bandwidths of FRB~20180916B bursts in Figure~\ref{fig:r3bws}, it is clear that as a population they tend to have narrow emission bandwidths. Moreover, it shows that the narrow bandwidths associated with bursts from this source are real, as they are often detected when the beam response as a function of frequency is flat, in the center of the beams. In Figure~\ref{fig:r3bws}, we converted the one-dimensional distribution into probability density functions for visualization, using kernel density estimation for data with heteroscedastic errors\footnote{The errors being heteroscedastic, as opposed to homoscedastic, means that not all data points have the same error associated with them.} \citep{dm08} as implemented in the CRAN package \texttt{decon}\footnote{\url{https://CRAN.R-project.org/package=decon}} \citep{ww11}.

\section{Fast radio burst archetypes}
\label{sec:archetypes}

In the observed sample of FRBs in the 400--800\,MHz octave at the 0.98304 ms time resolution of the real-time CHIME/FRB system, we identify -- by eye -- broadly four archetypes of FRB morphology:
\begin{enumerate}
    \item Broadband simple bursts comprised of one peak that can be reasonably well described by one Gaussian function in time, convolved with an exponential scattering tail if scattering is not negligible. Their spectra can be well described by a power-law function;
    \item Narrowband simple bursts -- with spectra that are more like Gaussians;
    \item Complex bursts comprised of multiple peaks with similar frequency extent, with one of the peaks sometimes being a much dimmer precursor or postcursor – they can be broadband or narrowband;
    \item Complex bursts comprised of multiple subbursts that drift downward in frequency as time progresses.
\end{enumerate}

\begin{figure*}
    \centering
    \includegraphics{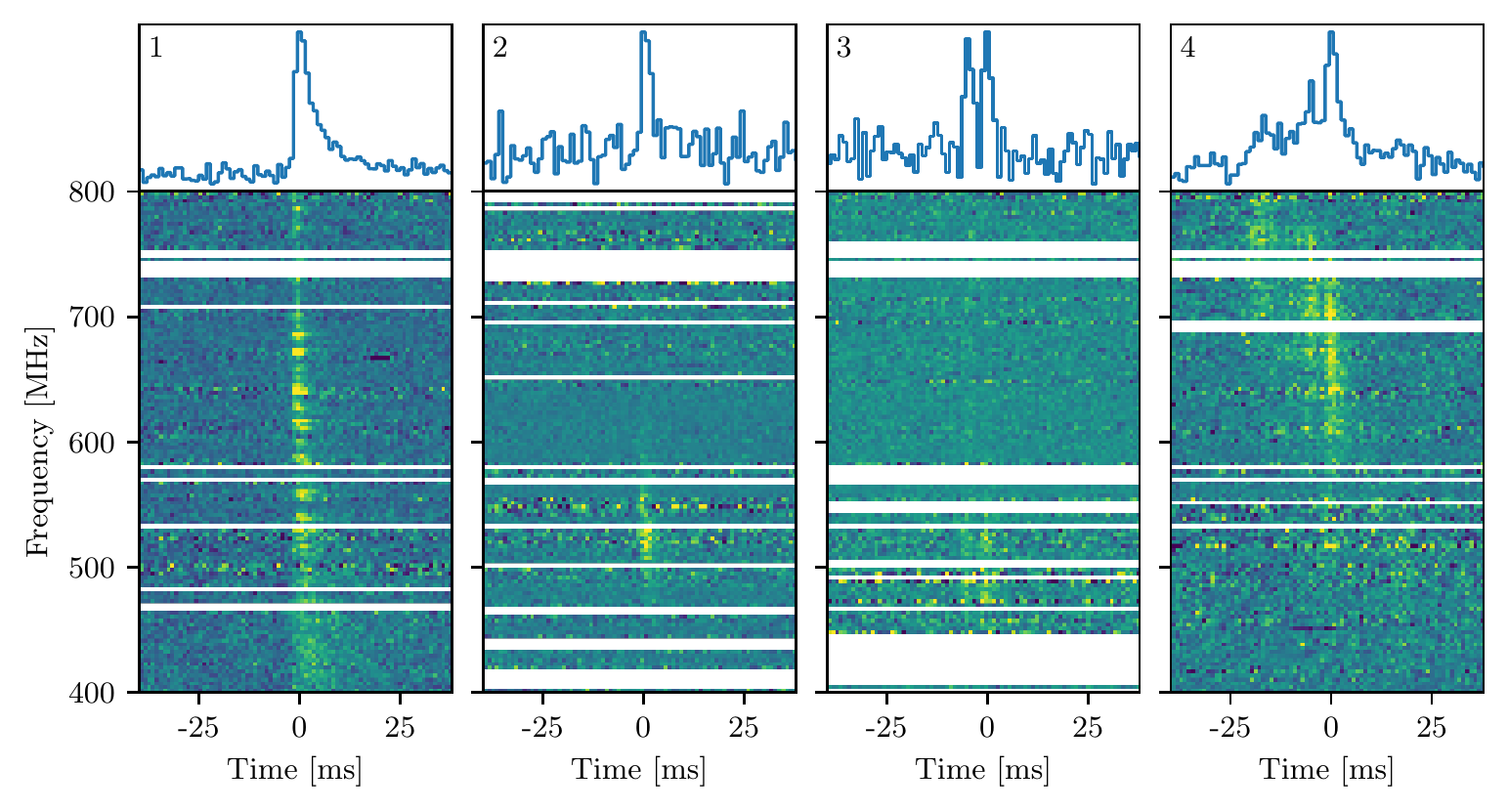}
    \caption{Dedispersed dynamic spectra of the four FRB morphology archetypes described in the text, as detected by CHIME/FRB. These are FRBs~20190527C, 20190515D, 20181117B and the 2019 August 10 burst of repeating FRB~20190117A. Knots of intensity in the burst spectra are instrumental in origin.}
    \label{fig:archetypes}
\end{figure*}

One example FRB of each archetype, as detected by CHIME/FRB, is shown in Figure~\ref{fig:archetypes}. In Catalog 1, the archetypes describe roughly 30\%, 60\%, 5\% and 5\% of the bursts, respectively.

There are multiple ways in which the instrumental response of the telescope can obscure the morphology of a burst as compared to how it arrived at the observatory: as described in \S\ref{sec:beameffects}, beam effects can suppress emission in parts of the band, making a burst appear narrower in frequency (type 1 $\rightarrow$ 2) and potentially concealing drifting subbursts (type 4 $\rightarrow$ 3). Bursts with structure below the time resolution or with low S/N will be unresolved (type 4 $\rightarrow$ 2). Furthermore, in the context of this paper, ``broadband'' is used to describe emission that extends roughly from 800\,MHz down to 400\,MHz; however, it does not necessarily imply that the burst emission extends over many GHz of bandwidth. Moreover, drift rates $\gtrsim 400$\,MHz ms$^{-1}$ (receiver bandwidth over time resolution) will not be identified as drifting, appearing as type 3 instead of 4 \citep[see also the discussion on drift rate measurements in][]{abb+19c}.

A broadband and singlepeaked morphology has so far never been clearly observed in a repeater burst detected by CHIME/FRB \citep[see dynamic spectra in][]{abb+19b,abb+19c,jcf+19,fab+20,aab+20} and downward-drifting subbursts seem ubiquitous among repeater bursts \citep[detected in at least 19 bursts from 13 different sources among the 58 bursts from 19 sources in][]{abb+19b,abb+19c,jcf+19,fab+20}.

\section{Repeaters versus nonrepeaters}
\label{sec:versus}

\begin{figure*}
\centering
\includegraphics{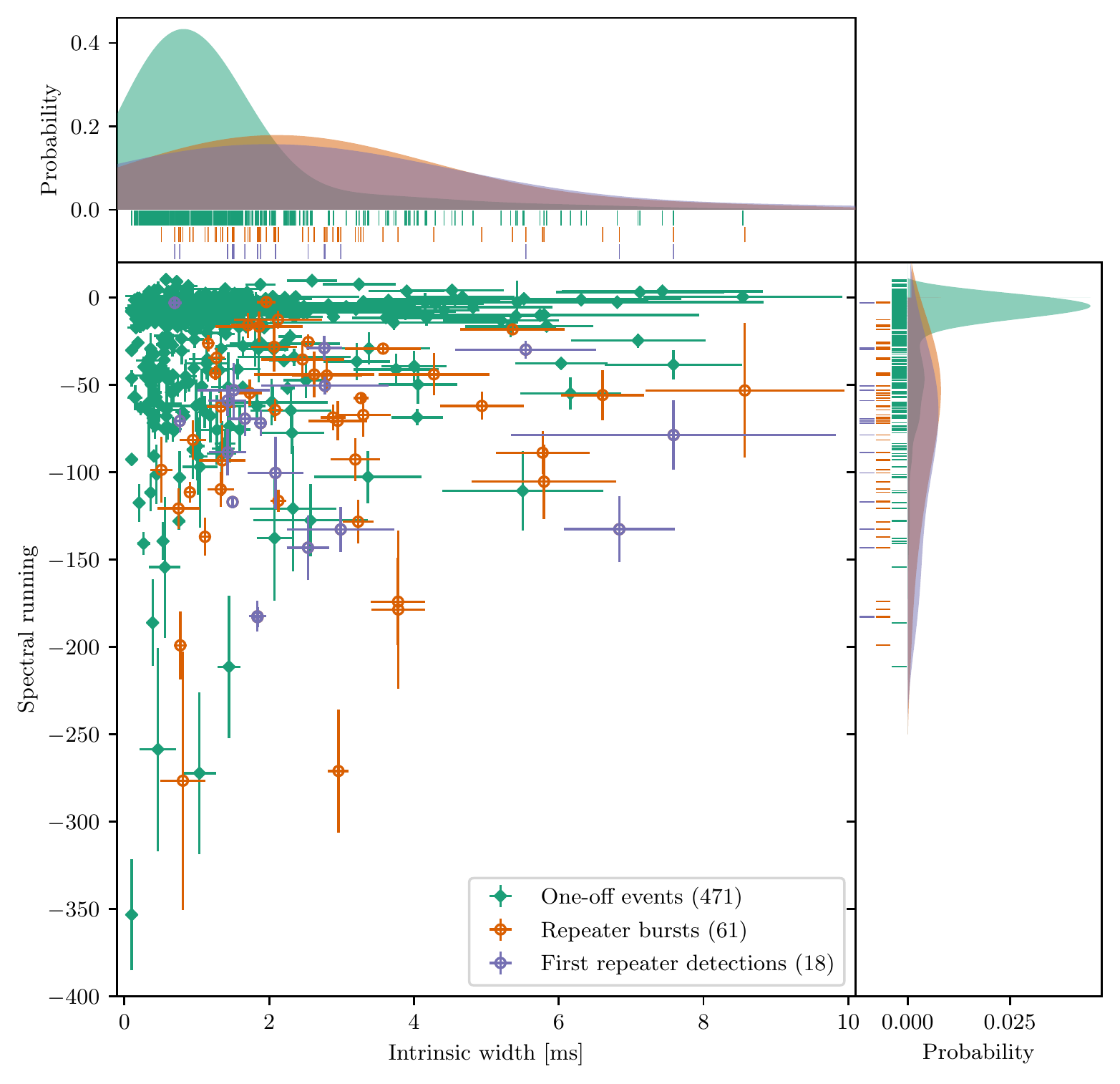}
\caption{Spectral runnings and intrinsic widths from model fits to the FRBs in Catalog 1, separated in one-off events (green diamonds), repeater bursts (orange open circles) and first repeater detections (purple open circles). Error bars represent $1\sigma$ uncertainties on the least squares model fits. On the top and on the right the 1D distributions (vertical/horizontal lines) and associated kernel density estimates (see \S\ref{sec:beameffects}) of the respective parameters.}
\label{fig:cat_width_running}
\end{figure*}

\begin{figure*}
\centering
\includegraphics{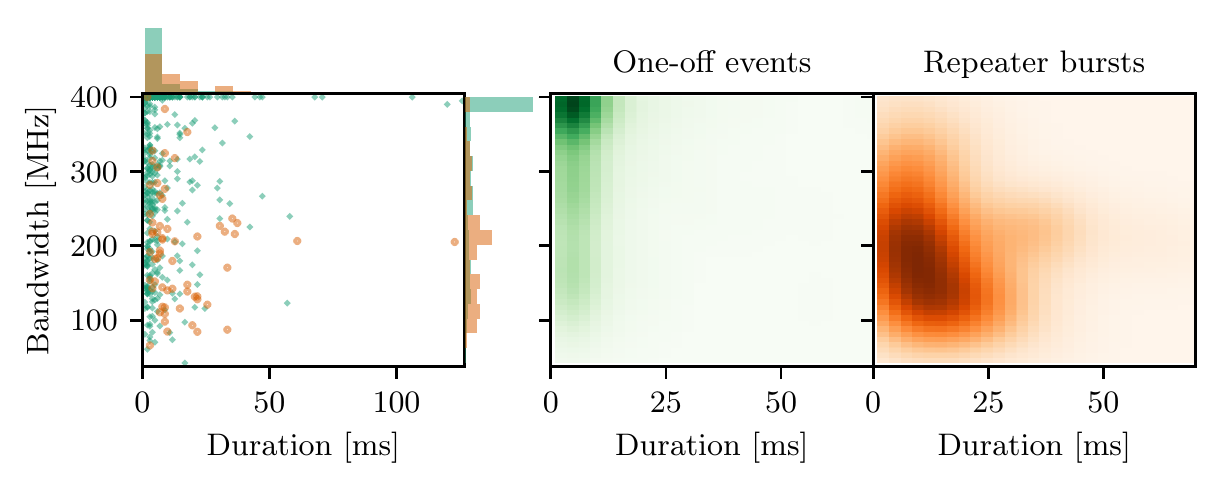}
\caption{Bandwidths and durations of the FRBs in Catalog 1 with normalized histograms on the sides. FRBs are separated into one-off events (green diamonds) and repeater bursts (orange open circles). The panels on the right show smoothed and normalized distributions of all one-off events (green) and all repeater bursts (orange), respectively. Only bursts with detection S/N $>$ 12 are included. Note that the one-off events may contain as-of-yet undiscovered repeaters sources (e.g., due to limited exposure or source activity). Note also that a subset of bursts have underestimated bandwidths due to detections away from the center of a formed beam (cf.~\S\ref{sec:beameffects}).}
\label{fig:cat_duration_bw}
\end{figure*}

In Figure~\ref{fig:cat_width_running}, we compare the morphology of FRBs that have not yet repeated with those of FRBs from repeater sources  using the fitted intrinsic widths and spectral runnings as a proxy for bandwidth. We consider the sample of all repeater bursts as well as the sample with only the first repeater detections, as the former is biased by the detection threshold for subsequent bursts from the same source being lower and by containing a disproportionate number of bursts from more prolific repeater sources. In Figure~\ref{fig:cat_duration_bw}, we compare the durations and bandwidths of FRBs. This is a slightly more intuitive comparison as the two values can be more easily estimated by eye from a dynamic spectrum.

By eye the parameters associated with one-off events and repeater bursts already appear to be distributed differently. Even though burst spectra are described by two parameters (the spectral index $\gamma$ and running $r$, or bandwidth and peak frequency), we compare the parameter distributions separately, as nonparametric statistical tests that ask whether two samples are drawn from the same probability distribution are ill-defined in more than one dimension \citep[e.g.,][]{fb12}.

We take the intrinsic widths, burst durations, spectral runnings and burst bandwidths and compare all one-off FRBs with the first detected bursts from repeaters, using the k-sample Anderson-Darling (AD) test \citep{ss87}, as implemented in the CRAN package \texttt{kSamples}.\footnote{\url{https://CRAN.R-project.org/package=kSamples}} We do not include a comparison of the spectral indices as this parameter is correlated with the spectral running and the running more directly distinguishes broadband ($r \sim 0$) and narrowband ($r < 0$) emission. To account for the uncertainties on the measurements of the intrinsic width and spectral running we Monte Carlo (MC) resample 10,000 times, each iteration drawing a measurement randomly from its Gaussian uncertainty region before recalculating the p-value from the AD test comparison. Across all MC iterations we calculate a 95\% confidence interval upper limit on the p-values. We repeat the analysis after imposing a detection S/N $\geq 12$ threshold for inclusion (below this threshold real events that were detected in the search might have been misclassified as noise or radio frequency interference by humans verifying candidate FRBs). The results of the analyses, compiled in Table~\ref{tab:ad}, show that among all four variables there exists a statistically significant observed difference between the two samples. It is thus clear that one-off events and repeater bursts differ strongly in average width and bandwidth.

\begin{table*}[t]
\begin{center}
\caption{One-dimensional statistical comparisons of one-off FRBs and repeater bursts, including all FRB subbursts (``individual'') or per-FRB averages of subburst parameters (``averaged''). Comparisons are done without and with a cutoff on S/N = 12, with the results for the samples with a S/N cutoff added in parentheses.}
\begin{tabular}{lllccccc}
\tableline \tableline
One-off events vs. & Subbursts & Sample size & \multicolumn{4}{c}{p-value} \\ \cline{4-7}
& & & Width\tablenotemark{a} & Duration & Spectral running\tablenotemark{a} & Bandwidth \\ \tableline
Repeater firsts & Individual & 503 (342) vs. 24 (16) & $10^{-3}$ ($10^{-3}$) & $10^{-7}$ ($10^{-8}$) & $10^{-12}$ ($10^{-12}$) & $10^{-6}$ ($10^{-5}$) \\
Repeater firsts & Averaged & 471 (314) vs. 18 (10) & $10^{-4}$ ($10^{-3}$) & $10^{-4}$ ($10^{-4}$) & $10^{-9}$ ($10^{-7}$) & $10^{-5}$ ($10^{-3}$) \\
Repeater bursts & Individual & 503 (342) vs. 93 (57) & $10^{-13}$ ($10^{-11}$) & $10^{-21}$ ($10^{-20}$) & $10^{-36}$ ($10^{-24}$) & $10^{-20}$ ($10^{-12}$) \\
Repeater bursts & Averaged & 471 (314) vs. 61 (33) & $10^{-12}$ ($10^{-9}$) & $10^{-10}$ ($10^{-10}$) & $10^{-24}$ ($10^{-16}$) & $10^{-13}$ ($10^{-8}$) \\
\tableline
\end{tabular}
\tablenotetext{\mathrm{a}}{95\% confidence upper limit from 10,000 Monte Carlo trials.}
\label{tab:ad}
\end{center}
\end{table*}

\section{Subburst separations}
\label{sec:subbursts}

For FRBs that have multiple components, the difficulty in asserting a single event (under a single ``envelope''), as opposed to multiple events close in time, grows larger as the subburst separation increases. Using the model fits from Catalog 1, we can test whether there is a bimodality in subburst separations (i.e., if the data support a threshold separation below which two subbursts are part of the same FRB envelope and above which two subbursts are instead two separate FRBs). At the same time, we can also investigate the separation between the frequency center of subsequent components. 

We calculate the time separation $\Delta t = \mathrm{TOA}_{i+1} - \mathrm{TOA}_i$ and the frequency separation $\Delta \nu_\mathrm{c} = \nu_{\mathrm{c},i+1} - \nu_{\mathrm{c},i}$ where TOA$_{i}$ is the time-of-arrival and $\nu_{\mathrm{c},i}$ the frequency center of component $i$. A negative $\Delta \nu_\mathrm{c}$ implies a downward drift. Note that there is an unquantified uncertainty in the peak frequency, especially for those subbursts that extend beyond the edges of the receiver bandwidth. As a separation is only defined for FRBs with two or more components, we only include those FRBs in this analysis. An FRB comprised of $n$ subbursts results in $n-1$ data points in this analysis.

\begin{figure}
    \centering
    \includegraphics{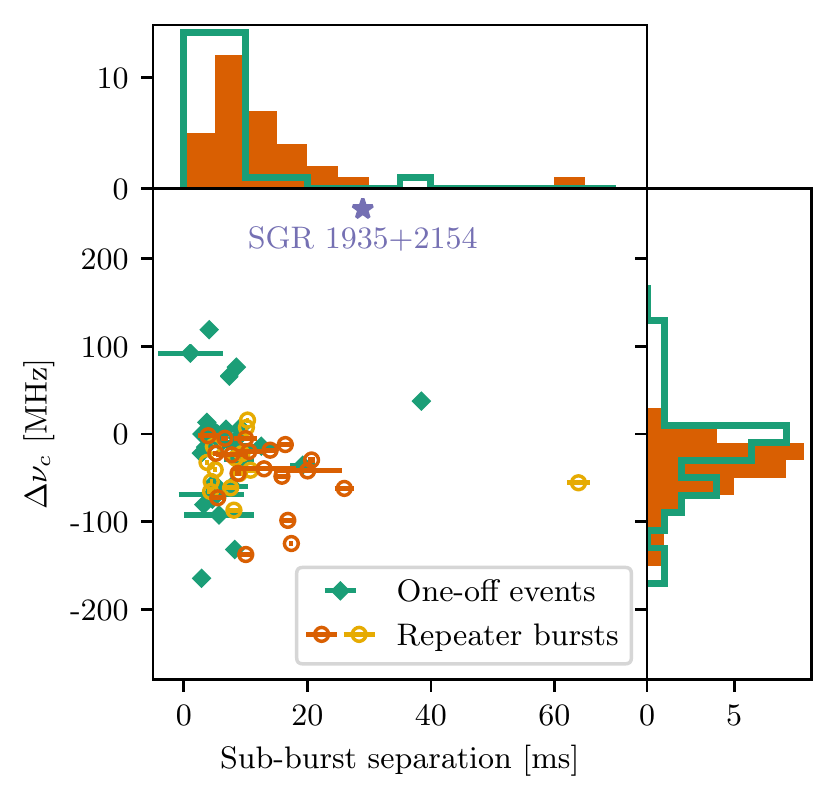}
    \caption{Subburst separations in time and frequency in Catalog 1 with bursts from apparent nonrepeaters (green diamonds) and repeater bursts (orange and yellow open circles). Note that only FRBs that have two or more subbursts are included. Of the 18 repeater bursts, 8 are from FRB~20180916B. Those 8 have been colored yellow in the central panel to avoid biasing the interpretation. The CHIME/FRB detection of a burst from SGR 1935+2154 (purple star) is shown for reference. On the top and on the right the 1D histograms of the respective parameters, including \emph{all} repeater burst pairs in orange.}
    \label{fig:subburst_seps}
\end{figure}

In Figure~\ref{fig:subburst_seps}, we compare all subburst separations. From the distribution of subburst separations in time there is no clear bimodality between short (here $\sim$ms) and long (here $\sim$tens of ms) separations in this sample, with only two bursts with separations $> 30$\,ms. The data seem consistent with being drawn from an exponential-like distribution with a long tail. This holds true for both one-off events and repeater bursts, with repeater bursts on average seeming to have larger subburst separations. However, note that the $\sim$1 ms time resolution is left-censoring the distribution. Considering the frequency separations, all repeater bursts have $\Delta \nu_\mathrm{c} \leq 0$, which means that they do not drift in frequency or that they drift downward in frequency.\footnote{Note though that bursts 24 and 25 from FRB~20180916B  (not included in Catalog 1) are separated in time by only about 60 ms and appear to have an upward drift of about 100\,MHz \citep{aab+20}.}

\begin{figure*}
    \centering
    \includegraphics{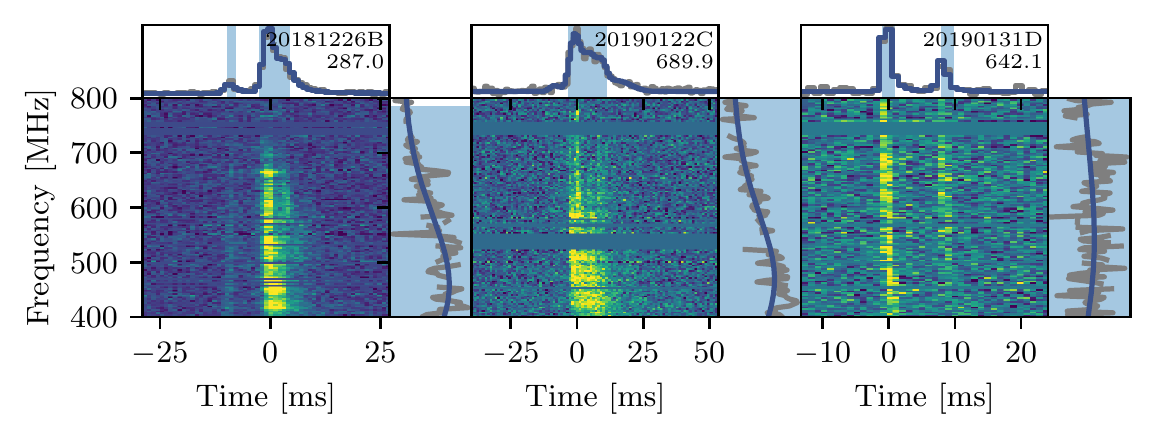}
    \caption{Dynamic spectra of FRBs 20181226B, 20190122C and 20190131D with three, six and two subbursts, respectively. These three FRBs show \emph{apparent} upward drifts between subbursts in Figure~\ref{fig:subburst_seps} (see \S\ref{sec:subbursts}). The band-averaged time series and time-averaged spectra are shown in the panels on the top and on the right, respectively, with the fitted models overplotted in a blue solid line. The burst durations and frequency extent are indicated by a blue shaded region. The burst names and DMs (in pc~cm$^{-3}$ units) are displayed in the panels with the time series. Data are from Catalog 1.}
    \label{fig:updrift_wfalls}
\end{figure*}

Some of the one-off events might turn out to be bursts from as-of-yet undiscovered repeater sources. The overlap of one-off events with the sample of repeater bursts may be due to undiscovered repeaters.\footnote{We select the best candidates based on burst morphology and exposure/burst rate in Appendix~\ref{sec:followup}.} Three one-off events have apparent upward drifts between some of their subbursts, but all three are also broadband bursts, where the determination of the center frequency within the band is ambiguous and the uncertainty on the center is difficult to quantify. The frequency separations for those bursts are thus likely consistent with being zero. The dynamic spectra of these FRBs, 20181226B, 20190122C and 20190131D, are shown in Figure~\ref{fig:updrift_wfalls} for reference.

For comparison, we have also added the CHIME/FRB-detected burst from SGR 1935+2154 to Figure~\ref{fig:subburst_seps}. The two observed components for this burst clearly have different spectra, with the first peaking at the bottom of the CHIME band and the second at the top. This results in an upward drift of the frequency centers. The two components are separated by $\sim$29 ms and could be independent bursts rather than subbursts of the same envelope \citep{abb+20}. This burst thus seems to be an outlier among the FRBs in Catalog 1 and its morphology is not standard for repeating FRBs. It should be noted, however, that this magnetar burst had exceptionally large observed flux density due to its proximity and that for a lot of more distant FRBs a second (or $n^\mathrm{th}$) component might be buried in the noise.

\section{Discussion}
\label{sec:discussion}

We find an observed difference in the burst morphologies of one-off FRBs and repeater bursts. It is unlikely that these differences are caused by instrumental effects, such as beam response towards the sky location of an FRB at the time of its detection or the time resolution of the data, because both populations are subject to nearly identical biases. Similarly, the CHIME/FRB selection function that Catalog 1 presents, based on injections of simulated events in the real-time system, affects both one-off event and repeater burst detections. The selection function shows that CHIME/FRB is only recovering a fraction of wide and highly scattered events, which implies that the differences in width/duration between the two populations might even be understated in the current sample.

Moreover, it is clear that many FRBs are not simply broadband, narrow impulses that are potentially scatter-broadened but that FRBs are often comprised of multiple subbursts. This becomes especially apparent at sub-ms time resolution \citep[e.g.,][]{ffb18,dds+20}. It can be difficult to disentangle intrinsic burst structure and propagation effects, especially for FRBs detected with limited time resolution or S/N \citep{gms+19}. This in turn, hinders the precise determination of DMs, scattering timescales and subburst separations.

Next we discuss the potential interpretations and implications of burst morphology for FRB models and applications, respectively. 

\subsection{Potential interpretations}

It is possible that a selection effect due to beamed emission causes the observed difference in bursts widths, assuming there exists a positive correlation between burst width and beaming angle \citep{cmg20}. A simulation  by \citet{cmg20} of this putative effect reproduces the difference in burst widths between one-off events and repeater bursts qualitatively. The model predicts a correlation between repeat rate and burst width that cannot be significantly detected with the currently published set of repeater sources. An explanation for the narrow bandwidths of repeater bursts in this framework is currently lacking, and not accounted for in any model to our knowledge. More generally (i.e., without necessarily invoking beaming), the two populations could be manifestations of two different regimes out of a continuum of source activity that is correlated with the burst morphology.

The differences might also be due to propagation effects. \citet{cwh+17} proposed plasma lensing as a way to explain the complex time/frequency patterns, which we now recognize as typical of the repeating FRB population, in the first discovered repeater, FRB~20121102A \citep[e.g.,][]{hss+19}. \citet{cwh+17} showed that impulsive radio waves passing through discrete AU-scale structures can experience strong frequency-dependent focusing that results in high-gain caustics that alter light curves and spectra. They suggested such structures could arise from one of three sources: a source-induced bow shock in a dense medium, filaments in a surrounding supernova remnant (SNR), or filaments in a surrounding \ion{H}{2} region. All these options tend to implicate environments like those in the disks of spiral galaxies or star-forming regions. However, if this were the correct interpretation for the time/frequency structure of repeaters, then it suggests apparent nonrepeaters must not be as commonly subject to plasma lensing, which argues for them having different environments. Yet, as shown in Catalog 1, repeaters and apparent nonrepeaters have DM distributions that are consistent with arising from the same underlying distribution. The same is true of their scattering time distributions. This suggests similar environments, at odds with the plasma lensing implications, unless, for example, the two classes represent only an age disparity, with repeaters somewhat younger (hence preferentially inside SNRs), but having largely similar galactic environments. Though there are small disparities between the host galaxies of the two classes, robust conclusions regarding their differences cannot yet be drawn due to the small number of repeating sources with host galaxy associations \citep{bsp+20,hpf+20}. We note that the strong preference for \emph{downward}-drifting subbursts in repeater bursts is problematic in the plasma lensing model as already noted by \citet{hss+19}; our results only exacerbate the problem.

A different possibility is that if repeater sources are preferentially located in binary systems \citep[cf. the periodic activity of FRBs~20180916B and 20121102A;][]{aab+20,rms+20,css+21} the intrabinary or circumbinary material might cause a propagation effect that leads to the wider bursts and narrower bandwidths. If so, then periodicities in repeat bursts should eventually be shown to be common. Plasma lensing has been observed in Galactic binary pulsars \citep{myc+18,brd19}.

Finally, it is possible that the observed differences are intrinsic, with one-off events and repeaters forming two distinct populations of FRBs produced by different progenitors and/or emission mechanisms. However, again, it would be puzzling that no differences have so far been seen between the observed DM and scattering distributions of the two populations if they have different progenitors (\citealt{fab+20}; Catalog 1). Perhaps one-off events and repeaters are produced through the same astrophysical formation channels, with the differences in burst rate and complexity of the FRB emission being due to a separation in age, magnetic field (strength and/or orientation) or some other parameter intrinsic to the source.

Some proposed models for FRBs predict narrowband emission: decelerating blast waves result in $\Delta \nu / \nu \sim 1$ emission \citep{mms19}, which is narrowband, but still necessitates some external effect to produce the $\Delta \nu / \nu \sim 0.1$ that is observed for repeater bursts. Superradiance in a molecular cloud triggered by a coherent pulse from, e.g., a pulsar, appears to be able to explain repeater burst bandwidths and spectral structure \citep{hrg+19}. These models, however, do not predict a difference between one-off events and repeater bursts. The interaction of hypothetical axion stars with neutron stars and the accretion disks of black holes has been proposed to lead to one-off events and repeating FRBs, respectively, with different center frequencies and bandwidths of emission between the two populations \citep{iwa20}.

Repeater bursts might be made up of multiple subbursts analogous to the main pulse of the Crab pulsar \citep[see, e.g.,][]{cw16}, with many of the subburst separations below the time resolution of the CHIME/FRB $\sim$1 ms data and most other detection systems. If some physical process causes a distribution of subburst separations that peaks below the time resolution of an instrument, it is most likely to detect burst separations close to the time resolution of that instrument, as can be observed in Figure~\ref{fig:subburst_seps}. Mapping out the precise wait-time distribution will help differentiate between FRB emission models and, e.g., the CHIME/FRB baseband data will help address this.

\subsection{FRB classification}

Our results show that FRBs can be probabilistically classified as either a one-off event or a repeater burst, based on only their burst morphologies. In Appendix~\ref{sec:followup} we present one-off events from Catalog 1 that are likely to yield repeat bursts, based on their morphology. As a next step, a predictive model can be built using a machine learning algorithm, to investigate how accurately this classification can be performed with the current data set. These kind of classifications can be made more precise in the future by not only considering the burst morphologies, but also, e.g., the brightnesses, distances, polarization information and host galaxy associations of the FRBs. A multiparameter comparison will also likely reveal if there exist more than two observed classes of FRBs. Concurrently, it might be possible to train a classifier on the FRB dynamic spectra directly, as has been successfully done to unambiguously separate short and long GRBs \citep{jss+20}. From expanding the CHIME/FRB sample of FRBs and a more quantitative understanding of the experiment's selection effects we will soon be able to determine if the observed classes of FRBs are distinct or instead various realizations of one continuum of FRB properties.

\subsection{Implications for FRBs as astrophysical tools}

FRBs have been proposed as probes for solving long-standing open questions in astrophysics, such as weighing the ``missing'' baryons in the intergalactic medium \citep[e.g.,][]{mcq14,mpm+20}, measuring the reionization history of the Universe \citep[e.g.,][]{ino04} and the optical depth of the cosmic microwave background \citep[e.g.,][]{fl16b}. These applications depend on measuring FRB DMs and redshifts. Even though we have shown that burst morphology limits the determination of DMs to $\sim$0.1 pc~cm$^{-3}$ precision in many cases\footnote{Though note that the achievable precision depends on the observing frequency and the resolution and sensitivity of the instrument used.}, this will not affect the use of FRBs for these applications, as the error budgets for these analyses are dominated by the uncertainty on host galaxy contributions to DM$_\mathrm{FRB}$ that are of order $100$ pc~cm$^{-3}$.

The progenitors and local environments of repeating FRBs can be probed by DM evolution of the bursts. FRBs are often theorized to live in dense environments, like pulsar wind nebulae or supernova remnants, especially in models in which the FRB source is a young neutron star. In those cases, the rapidly changing environment can introduce significant DM changes up to a few tens of pc~cm$^{-3}$ over the first few years after the supernova \citep{pg18}, that can be probed with $\sim$0.1 pc~cm$^{-3}$ precision. It is also possible to probe the kind of orbital DM variations due to changes in the line-of-sight electron density observed for binary Galactic pulsars with B (up to 4 pc~cm$^{-3}$) or Be (up to 11 pc~cm$^{-3}$) type companions \citep{msk+12,wjm04} but not the $10^{-3}$ pc~cm$^{-3}$ variations for pulsars with brown dwarf companions \citep{myc+18}. It has furthermore been calculated that the propagation of luminous FRBs (isotropic energy of $10^{45}$ erg s$^{-1}$) can accelerate electrons in the local environment ($< 1$~pc) of an FRB source, which can lead to burst-to-burst DM variations up to $10^{-2}$ pc~cm$^{-3}$ depending on the wave strength of each burst \citep{lw20}. These kind of measurements can only be made with better DM precision at $\sim\mu$s time resolution \citep{cms+20}.

FRBs can potentially be observably gravitationally lensed and probe dark matter distributions in the Universe \citep[e.g.,][]{mkdk16,sme+20}. Whether wave effects are directly visible in the burst morphology of an FRB depends on the Fresnel scale of the lensing scenario, which in turn depends on the physical parameters of the lens and the observing frequency. In the infinite frequency (``geometric optics'') limit (or if FRB images decohere due to the presence of a scattering screen in the FRB's line-of-sight), coherent effects due to the wave nature of light are unimportant. In this case two lensed copies will appear as two bursts with similar morphologies and polarimetric properties, but with different magnifications. In the finite high frequency (``eikonal'') limit, interference patterns typical of a double-slit experiment may be seen. In the low frequency (``wave optics'') limit, effects of diffraction may be observable as well \citep{jfp+20,jt20}.

A first hint that two FRBs are lensed copies of each other could thus be similar burst morphology and sky location. An unambiguous detection of lensing can be made when the electric fields of two FRBs correlate. In the geometric optics limit, the detection of a lensed event could be claimed based on solely the similar burst morphology and polarization (on top of similar sky location), in the absence of strongly chromatic effects in the vicinity of the lens. The morphological diversity among especially repeater bursts could make the identification of potentially lensed events easier (i.e., it would be more difficult to assert two events are the same if all events were identical). 

\section{Conclusions}
\label{sec:conclusion}

We have considered the burst morphologies of all FRBs in the first CHIME/FRB catalog using simple model fitting to their dynamic spectra. We have compared the widths and bandwidths of one-off FRBs with those of repeater bursts. There is a statistically significant observed difference between the two, with repeater bursts on average having a longer duration and being narrower in bandwidth.

This difference could be due to a beaming or propagation effect, or it can be intrinsic to the populations. We have also considered subburst separations in the first CHIME/FRB catalog. We find no clear bimodality in separation between subbursts that clearly fall under one envelope and those without a ``bridge'' in emission between subbursts in this sample.

Looking forward to a second CHIME/FRB catalog, better burst-fitting techniques will lead to more accurate model fits and thus a better morphological population comparison. For example, the beam model and sky position uncertainty regions could be taken into account to get a better handle on the bandpass response and hence the burst bandwidths. FRB subburst separations can be investigated further by using CHIME/FRB baseband data that are saved to disk for about a hundred FRBs that are part of the first CHIME/FRB catalog. These data, with a time resolution of 2.56\,$\mu$s, can resolve a lot of the structure in FRBs that remained unresolved in this analysis. It will potentially reveal a plethora of burst structures below a ms. Additionally, the baseband data can be phased up towards the best known sky position of an FRB, which results in data that are unaffected by the synthesized beam (though not the telescope primary beam) that causes much of the instrumental response that affects the intensity data \citep{mmm+20}.

Many one-off FRBs span the full CHIME band, but it is not clear how broadband their spectra really are. Future very-wide-band detections of one-off FRBs or detections with multiple facilities at the same time will reveal how far the bursts' instantaneous spectra reach. 

Finally, we see two possible avenues for solving the conundrum of the significant difference in repeater versus apparent nonrepeater burst properties, in spite of the two classes' similarity in DM and scattering time distributions.  One may be via comparative studies of low-DM sources, which, due to their proximity, could eventually reveal differences in multi-wavelength properties which could solidify distinct natures. Alternatively, a large number of sub-arcsecond FRB localizations could eventually help pinpoint locations within host galaxies and show differences between repeating sources and one-off events that will support different source classes.  This should be enabled by the upcoming construction of CHIME/FRB Outrigger telescopes at continental baselines from CHIME \citep{lmm+21}.

\begin{acknowledgements}
We acknowledge that CHIME is located on the traditional, ancestral, and unceded territory of the Syilx/Okanagan people.

We thank Emily Petroff and the anonymous referee for comments that have improved the quality of this manuscript.

We thank the Dominion Radio Astrophysical Observatory, operated by the National Research Council Canada, for gracious hospitality and expertise. CHIME is funded by a grant from the Canada Foundation for Innovation (CFI) 2012 Leading Edge Fund (Project 31170) and by contributions from the provinces of British Columbia, Qu\'ebec and Ontario. The CHIME/FRB Project is funded by a grant from the CFI 2015 Innovation Fund (Project 33213) and by contributions from the provinces of British Columbia and Qu\'ebec, and by the Dunlap Institute for Astronomy and Astrophysics at the University of Toronto. Additional support was provided by the Canadian Institute for Advanced Research (CIFAR), McGill University and the McGill Space Institute via the Trottier Family Foundation, and the University of British Columbia. The Dunlap Institute is funded through an endowment established by the David Dunlap family and the University of Toronto. Research at Perimeter Institute is supported by the Government of Canada through Industry Canada and by the Province of Ontario through the Ministry of Research \& Innovation. The National Radio Astronomy Observatory is a facility of the National Science Foundation (NSF) operated under cooperative agreement by Associated Universities, Inc. FRB research at UBC is supported by an NSERC Discovery Grant and by the Canadian Institute for Advanced Research. The CHIME/FRB baseband system is funded in part by a CFI John R. Evans Leaders Fund award to IHS.

D.C.G. is supported by the John I. Watters research fellowship. V.M.K. holds the Lorne Trottier Chair in Astrophysics \& Cosmology, a Distinguished James McGill Professorship and receives support from an NSERC Discovery Grant (RGPIN 228738-13) and Gerhard Herzberg Award, from an R.~Howard Webster Foundation Fellowship from CIFAR, and from the FRQNT CRAQ. S.M.R. is a CIFAR Fellow. P.S. is a Dunlap Fellow and an NSERC Postdoctoral Fellow. K. B. is supported by an NSF grant (2006548). M.B. and P.C. are supported by FRQNT Doctoral Research Awards. B.M.G. acknowledges the support of the Natural Sciences and Engineering Research Council of Canada (NSERC) through grant RGPIN-2015-05948, and of the Canada Research Chairs program. C.L. was supported by the U.S. Department of Defense (DoD) through the National Defense Science \& Engineering Graduate Fellowship (NDSEG) Program. K.W.M. is supported by an NSF Grant (2008031). J.M.P is a Kavli Fellow. D.M. is a Banting Fellow. D.M. is a Banting Fellow. K.S. is supported by the NSF Graduate Research Fellowship Program.
\end{acknowledgements}

\facilities{CHIME/FRB}

\software{Matplotlib \citep{matplotlib}, NumPy \citep{numpy}, rpy2 (\url{https://rpy2.github.io/}), SciPy \citep{scipy}, scikit-learn \citep{scikit-learn}}

\bibliographystyle{aasjournal}

\bibliography{frbrefs}

\appendix

\section{Candidate repeater sources}
\label{sec:followup}

\begin{figure*}[b]
    \centering
    \includegraphics{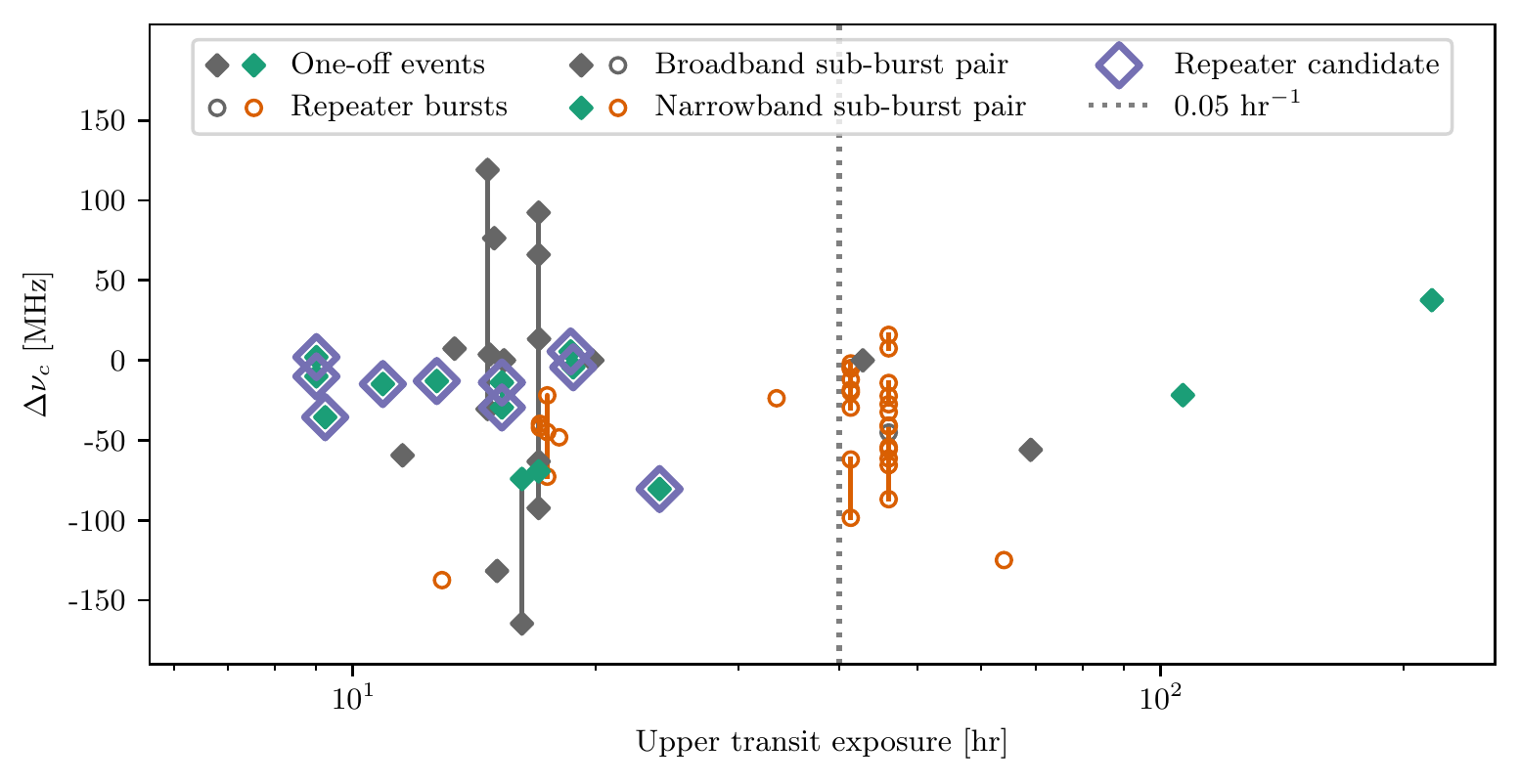}
    \caption{Subburst pair separation in frequency and source exposure (in the CHIME/FRB upper transit). Pairs of subbursts from one-off events (green/gray filled diamonds) and repeater bursts (orange/gray open circles) are shown. Multiple pairs of subbursts associated with the same event are linked with vertical lines. Non-gray markers indicate narrowband pairs (bandwidth $<$ 300\,MHz). A source with a Poisson burst rate of 0.05 hr$^{-1}$ (gray dotted line) -- roughly the lowest observed rate for a CHIME/FRB repeater \citep{fab+20} -- is expected to burst at least twice (i.e., is observed to repeat) in 40 hours. A one-off event with $>$ 40 hours of exposure would thus have been detected as a repeater if the source would be as prolific as the known repeaters. Repeater candidates are identified with open purple diamonds.}
    \label{fig:dfs_exp}
\end{figure*}

As we have established that narrowband emission and downward-drifting subbursts are ubiquitous among repeater bursts, we select candidate repeater sources among the sample of one-off events in Catalog 1 that also show these characteristics and that have had relatively low exposure. These sources can be good candidates for follow-up observations. To deem an event a candidate repeater source, we require the subburst drift rate $\Delta\nu_\mathrm{c} < 30$\,MHz ms$^{-1}$ (in turn, this requires the number of subbursts to be $\geq$ 2), the bandwidth of each subburst $<$ 300\,MHz and the exposure to the source in CHIME/FRB's upper transit $<$ 40 hr \citep[which allows for the burst rate $\gtrsim$ 0.05 hr$^{-1}$ -- roughly the lowest observed rate for a CHIME/FRB repeater;][]{fab+20}. We find eight candidates based on these criteria. Although in Catalog 1 we only analyze the highest S/N beam detection of each event, we inspected the intensity data for the surrounding beams for each candidate repeater and we exclude two candidates for which we see evidence for more broadband emission by comparing multiple beams (see \S\ref{sec:beameffects}). In Figure~\ref{fig:dfs_exp} we compare subburst separation and CHIME/FRB source exposure, as well as subburst bandwidths, for pairs of subbursts in Catalog 1. Table~\ref{tab:followup} lists the six one-off events in Catalog 1 that are potential repeater sources based on the aforementioned criteria.

\begin{table}[b]
\begin{center}
\caption{One-off events in Catalog 1 that are good candidates for follow-up observations to detect repeat bursts. Parameters with their uncertainties (68\%; in parentheses) are from Catalog 1.}
\begin{tabular}{lccccc}
\tableline \tableline
FRB & R.A. & Decl. & DM & Exposure \\
& [$^\circ$] & [$^\circ$] & [pc~cm$^{-3}$] & [hr] \\ \tableline
20181125A & 147.9(2) & 33.9(4) & 272.19(2) & 9(9) \\
20190308C & 188.36(3) & 44.4(2) & 500.52(3) & 19(7) \\ 
20190527A & 12.5(2) & 7.99(7) & 584.6(2) & 9(7) \\
20190422A & 48.6(2) & 35.2(2) & 452.30(1) & 15(8) \\
20190423A & 179.7(2) & 55.3(2) & 242.6456(6) & 24(9) \\ 
20190601C & 88.5(2) & 28.47(6) & 424.066(6) & 11(7) \\ 
\tableline
\end{tabular}
\label{tab:followup}
\end{center}
\end{table}

\end{document}